\documentclass[runningheads]{llncs}
\usepackage{graphicx}
\usepackage{hyperref}
\usepackage{booktabs}
\usepackage[backend=biber,style=numeric, giveninits=true]{biblatex} 

\AtBeginBibliography{\footnotesize}

\usepackage{fancyhdr}

\begin{document}
%
%
\title{A dataset and classification model for Malay, Hindi, Tamil and Chinese music}
%
\author{Fajilatun Nahar\inst{1} \and Kat Agres\inst{2} \and Balamurali BT\inst{1} 
\and
Dorien Herremans\inst{1}}
\authorrunning{F. Nahar et al.}
%

\institute{Singapore University of Technology and Design, Singapore 
\email{fajilatun\_nahar@mymail.sutd.edu.sg}
\and
National University of Singapore, Singapore
}

\maketitle              
\begin{abstract}
In this paper we present a new dataset, with musical excepts from the three main ethnic groups in Singapore: Chinese, Malay and Indian (both Hindi and Tamil). We use this new dataset to train different classification models to distinguish the origin of the music in terms of these ethnic groups. The classification models were optimized by exploring the use of different musical features as the input. Both high level features, i.e., musically meaningful features, as well as low level
 features, i.e., spectrogram based features, were extracted from the audio files so as to optimize the performance of the different classification models.

\end{abstract}
\section{Introduction}

Singapore is a cultural melting pot, with a majority of Chinese, Malay and Indian individuals. It is thus no surprise that Singaporean music is influenced by several different ethnical groups. The earliest form of music in Singapore was traditional Malay music \cite{perera2010music}, which came from the original settlers of Singapore. They are now the second largest ethnic group in Singapore \cite{govwebsite}. Then came the Portuguese influence from the colonial occupation, followed by Chinese and Indian music from the immigrants of those countries \cite{perera2010music}. Decades of rich political and cultural history of Singapore has established the current tastes and genres of music in Singapore \cite{kong1999invention}. In this paper, we create a dataset of music fragments of the three largest ethnical influences in Singapore, namely, Chinese, Malay, and Indian. This allows us to develop machine learning models that can estimate the probability of a song belonging to a certain ethnical group. In future research, these newly developed models will be useful to analyse typical Singaporean songs such as the National Day Songs. 


Over the last decade, significant strides have been made regarding audio classification models for mood/emotion \cite{laurier2008multimodal, cheuk2020emotion,patra2018multimodal}, genre \cite{tzanetakis2002musical, correa2016survey}, hit prediction \cite{herremans2014dance} and other topics. Most related to this research is the work on folk tune classification \cite{conklin2013multiple, chai2001folk}. Here, we focus on contemporary music from different Asian ethnical groups.

In the next section, we will discuss the dataset that we have gathered, followed by the extracted features and developed classification models in Section 3. The performance of our classifiers is presented in Section 4 and the final conclusion is presented in Section 5.

\section{Dataset creation}
We used the Spotify API\footnote{\url{https://developer.spotify.com/}} to retrieve a list of songs for each of our ethnical groups. The songs were manually curated by the first author, using search terms in the Spotify API. General search terms like `Hindi songs', `Chinese songs', `Malay songs' and `Tamil songs' were used, as well as names of popular singers of that specific ethnical group. A total of 15,725 songs were downloaded using the API, of which 3,146 were Chinese songs, 507 Malay songs, 6,729 Hindi songs, and 5,343 Tamil songs. We downloaded the first 30 seconds of the selected songs, some of which are instrumental songs, some contain only vocals, and some are a mix of both. Of these songs, a total of 260 low-level features and 98 high-level features were extracted using Essentia \cite{bogdanov2013essentia} and OpenSMILE \cite{eyben2015opensmile} respectively. For high-level features, six of the features were categorical features, so those features were one hot encoded, which increased the feature space to a total of 127 features. For low-level features, temporal data was collected in 0.5 seconds frames, totalling 58 frames per song. \textbf{These features were averaged for each song}. A detailed description of the features and the dataset itself is available online \footnote{\url{http://dorienherremans.com/sgmusic}}

Given the large number of extracted features, we do a preliminary exploration of which feature subset is most efficient  in the next section.

\section{Classification models}

There exist many types of classification algorithms that have shown to be effective for audio classification tasks. It is not the intention of this investigation to develop novel architectures or implement complex neural network structures. Instead, we focus on a very influential factor: input features. As per \cite{balamurali2019toward}, features greatly influence the performance of models. In this research, we hence focus on comparing different input representations (both high and low level music features) in basic, fast, and efficient machine learning models that have proven their efficacy in audio classification: logistic regression, k-nearest neighbours (k-NN), support vector machines (SVM) (with Grid search), and random forest.

The dataset was split into a training and test set with a ratio of 80:20. These models were tested using different feature subsets. These subsets can contain different types of features, and might include a feature selection mechanism, as described in Table \ref{tab:tab1}. This analysis reveals the most effective features for ethnical origin classification on our new dataset.   

Two feature selection methods were implemented: 1) A one-way ANOVA test is used to perform the filter method \cite{guyon2003introduction} where the the $p$-value is calculated for each feature. Features with a $p$-value of less than 0.05 are taken into consideration for further analysis. 2) The other technique is the wrapper method \cite{maldonado2009wrapper}, where backward elimination was performed by taking subsets of the features to create models using logistic regression. The accuracy of this model was examined and, using an iterative procedure, features were removed. The feature selection process will be stopped when the classifier delivers the best performance.




\section{Experiments and results}


We set up a preliminary experiment to analyse the influence of different feature representations on the classifier performance. We explored different combinations of high/low level features, with or without feature selection, thus forming Subsets of our data. 
We should note that these subsets are imbalanced, hence we include the class weighted AUC in the results in Table \ref{tab:tab1}.

\begin{table}
\scriptsize
 \caption{Subset description and model results}
    \label{tab:tab1}
    \centering
    
    \begin{tabular}{lclllcc}
    \toprule
        Subset & No of features & Feature type & Feature selection method & Best Model & AUC & Accuracy \\
    \midrule  
    1 & 260 & low-level & NA & SVM & 0.94 & 0.79 \\
    2 & 127 & high-level & NA & RF & 0.88 & 0.70 \\
    3 & 387 & high \& low & NA & SVM & 0.94 & 0.79 \\
    4 & 1,820 & low-level & NA & SVM & 0.93 & 0.77 \\
    5 & 111 & low-level & wrapper & SVM & 0.94 & 0.80 \\
    6 & 82 & high-level & wrapper & RF & 0.88 & 0.70 \\
    7 & 182 & low-level & filter & SVM & \textbf{0.95} & \textbf{0.81} \\
    8 & 67 & high-level & filter & RF & 0.88 & 0.69 \\
    9 & 92 & low-level & filter+wrapper & SVM & \textbf{0.95} & \textbf{0.81} \\
    10 & 49 & high-level & filter+wrapper & RF & 0.86 & 0.69 \\
    \bottomrule
    \end{tabular}
   
\end{table}


\begin{figure}[htp]
  \centering
   
  {\includegraphics[width=0.35\textwidth]{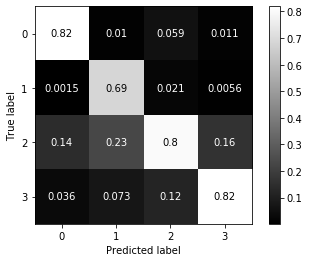}
  \label{fig:conf1}}
  {\includegraphics[width=0.35\textwidth]{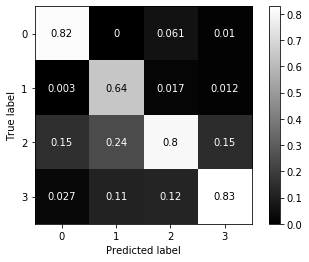}
  \label{fig:conf2}}
  \caption{Confusion matrices of two best performing models; left: Subset 9; right: Subset 7. Label 0 is Chinese; 1 is Malay; 2 is Hindi; 3 is Tamil.}
  \label{fig:confusion-matrix2}
\end{figure}

The SVM models using Subset 7 and 9 yielded the best AUC score of 95\% and an accuracy of 81\% on the test data. Both of these subsets contain only low-level features, and were reduced using feature selection methods.  The confusion matrices in Fig. \ref{fig:confusion-matrix2} also reveal a very similar performance of these two models. When comparing these two best performing models, we can conclude that Subset 9 is the more desired representation, because it contains less features, and as a result the training time is faster.

\section{Conclusions and future work}

We have gathered a dataset of $30s$ musical fragments together with 98 high and 260 low level musical features from four different ethnical origins. We have used this data to train relatively well performing classification algorithms. In an experiment, these classifiers perform best when using low-level audio features with feature selection as the input. In future research, we aim to further expand and visualise the songs of our dataset and make the models more robust, after which we can use them to explore the ethnical origin/influence of typical Singaporean music such as the National Day Songs.

 \printbibliography

@article{herremans2014dance,
  title={Dance hit song prediction},
  author={Herremans, Dorien and Martens, David and S{\"o}rensen, Kenneth},
  journal={Journal of New Music Research},
  volume={43},
  number={3},
  pages={291--302},
  year={2014},
  publisher={Taylor \& Francis}
}

@article{conklin2013multiple,
  title={Multiple viewpoint systems for music classification},
  author={Conklin, Darrell},
  journal={Journal of New Music Research},
  volume={42},
  number={1},
  pages={19--26},
  year={2013},
  publisher={Taylor \& Francis}
}

@inproceedings{chai2001folk,
  title={Folk music classification using hidden Markov models},
  author={Chai, Wei and Vercoe, Barry},
  booktitle={Proc. of Int. conf. on artificial intelligence},
  volume={6},
  number={6.4},
  year={2001},
  organization={sn}
}

@article{tzanetakis2002musical,
  title={Musical genre classification of audio signals},
  author={Tzanetakis, George and Cook, Perry},
  journal={IEEE Tran. on speech and audio processing},
  volume={10},
  number={5},
  pages={293--302},
  year={2002},
  publisher={IEEE}
}

@article{correa2016survey,
  title={A survey on symbolic data-based music genre classification},
  author={Corr{\^e}a, D{\'e}bora C and Rodrigues, Francisco Ap},
  journal={Expert Syst. Appl.},
  volume={60},
  pages={190--210},
  year={2016},
  publisher={Elsevier}
}

@inproceedings {cheuk2020emotion,
	title = {The impact of Audio input representations on neural network based music transcription},
	booktitle = {Proc. of the Int. Joint conf. on Neural Networks (IJCNN)},
	year = {2020},
	address = {Glasgow},
	author = {K.W. Cheuk and K. Agres and D. Herremans}
}

@inproceedings{laurier2008multimodal,
  title={Multimodal music mood classification using audio and lyrics},
  author={Laurier, Cyril and Grivolla, Jens and Herrera, Perfecto},
  booktitle={Int. conf. on Machine Learning \& Appl.},
  pages={688--693},
  year={2008},
  organization={IEEE}
}

@article{balamurali2019toward,
  title={Toward robust audio spoofing detection: A detailed comparison of traditional and learned features},
  author={Balamurali, BT and Lin, Kinwah Edward and Lui, Simon and Chen, Jer-Ming and Herremans, Dorien},
  journal={IEEE Access},
  volume={7},
  pages={84229--84241},
  year={2019},
  publisher={IEEE}
}

@article{perera2010music,
  title={Music in Singapore: From the 1920s to the 2000s},
  author={Perera, Loretta Marie and Perera, Audrey},
  journal={National Library Board, Singapore},
  year={2010}
}

@article{eyben2015opensmile,
  title={openSMILE:) The Munich open-source large-scale multimedia feature extractor},
  author={Eyben, Florian and Schuller, Bj{\"o}rn},
  journal={ACM SIGMultimedia Records},
  volume={6},
  number={4},
  pages={4--13},
  year={2015},
}

@inproceedings{bogdanov2013essentia,
  title={ESSENTIA: an open-source library for sound and music analysis},
  author={Bogdanov, Dmitry and Wack, Nicolas and G{\'o}mez, Emilia and Gulati, Sankalp and Herrera, Perfecto and Mayor, Oscar and Roma, Gerard and Salamon, Justin and Zapata, Jos{\'e} and Serra, Xavier},
  booktitle={Proc. of the 21st ACM Int. conf. on Multimedia},
  pages={855--858},
  year={2013}
}

@article{kong1999invention,
  title={The invention of heritage: popular music in Singapore},
  author={Kong, Lily},
  journal={Asian Studies Review},
  volume={23},
  number={1},
  pages={1--25},
  year={1999},
  publisher={Taylor \& Francis}
}

@article{patra2018multimodal,
  title={Multimodal mood classification of Hindi and Western songs},
  author={Patra, Braja Gopal and Das, Dipankar and Bandyopadhyay, Sivaji},
  journal={J. Intell. Inf. Syst.},
  volume={51},
  number={3},
  pages={579--596},
  year={2018},
  publisher={Springer}
}

@article{guyon2003introduction,
  title={An introduction to variable and feature selection},
  author={Guyon, Isabelle and Elisseeff, Andr{\'e}},
  journal={Journal of machine learning research},
  volume={3},
  number={Mar},
  pages={1157--1182},
  year={2003}
}

@article{maldonado2009wrapper,
  title={A wrapper method for feature selection using support vector machines},
  author={Maldonado, Sebasti{\'a}n and Weber, Richard},
  journal={Information Sciences},
  volume={179},
  number={13},
  pages={2208--2217},
  year={2009},
  publisher={Elsevier}
}

@online{govwebsite,
author = "",
    title = "What are the racial proportions among Singapore citizens?",
    url  = " https://www.gov.sg/article/what-are-the-racial-proportions-among-singapore-citizens",
    addendum = "(accessed: 6.08.2020)"
}

\end{document}